\documentclass{article}

\PassOptionsToPackage{numbers, compress}{natbib}
\usepackage[main, final]{neurips_2026}

\usepackage[utf8]{inputenc} 
\usepackage[T1]{fontenc}    
\usepackage{hyperref}       
\usepackage{url}            
\usepackage{booktabs}       
\usepackage{amsfonts}       
\usepackage{nicefrac}       
\usepackage{microtype}      
\usepackage{xcolor}         
\usepackage{amsmath}
\usepackage{amssymb}
\usepackage{graphicx}       

\title{StyleStream 2.0: Fast and Controllable Streaming Voice Style Conversion}

\author{%
  Yisi Liu$^{1}$,\enskip
  Nicholas Lee$^{1}$,\enskip 
  Gopala K. Anumanchipalli$^{1}$\\
 $^{1}$University of California, Berkeley
}

\begin{document}

\maketitle

\begin{abstract}
  Voice style conversion (VSC) aims to transform a source utterance to match the timbre, accent, and emotion of a target voice while preserving linguistic content.
  StyleStream 1.0 introduced the first streamable zero-shot VSC system with state-of-the-art conversion quality, but remained limited in two ways: approximately 1s end-to-end latency on consumer hardware, and the lack of a text-based interface for designing or editing target voices.
  We present StyleStream 2.0, a fast and controllable streaming VSC framework that addresses these limitations.
  To reduce latency, we train a few-step pixel MeanFlow model and further fine-tune it with autoregressive feedback inspired by Self Forcing, improving robustness under streaming inference.
  This reduces end-to-end latency to 520~ms on a consumer GPU, a 2x speedup over StyleStream 1.0, while maintaining conversion quality.
  To enable controllable style manipulation, we remove the mel-spectrogram context and instead route target style information through a compact style embedding.
  In this embedding space, we train a unified text-conditioned flow matching model that supports both text-based voice design, which maps natural language prompts to voice styles, and instruction-based voice editing, which modifies a specific attribute of a source utterance while preserving the rest.
  Experiments show that StyleStream 2.0 achieves the strongest target style fidelity on VSC, voice design, and voice editing, while remaining competitive on intelligibility and non-edited attribute preservation. Samples can be found here: \href{https://berkeley-speech-group.github.io/StyleStream2.0/}{https://berkeley-speech-group.github.io/StyleStream2.0}.
\end{abstract}
\section{Introduction}

Voice style conversion (VSC) modifies how an utterance is spoken while preserving what is said.
Following StyleStream 1.0 \cite{liu2026stylestream}, we define voice style as the joint combination of timbre, accent, and emotion.
A large body of voice conversion work focuses on timbre conversion, aiming to change speaker identity while preserving content and other paralinguistic factors \cite{qian2019autovc, li2023freevc, wang2026diffvc, choi2021neural, baas2023voice, liu2025rt, yang2024streamvc}.
Separate lines of work address accent conversion \cite{jia2024convert, bai2025accent, jia2023zero} and emotional voice conversion \cite{zhou2022emotion, pan2025clapfm, chou2025zsdevc} as standalone tasks.
More recent systems have begun to address holistic style transfer, jointly converting timbre, accent, and emotion from a single reference utterance \cite{zhang2025vevo, amphion_v0.2, zhang2025vevo2, anastassiou2024voiceshop, liu2026stylestream}. 
Among them, StyleStream 1.0 \cite{liu2026stylestream} is the first to demonstrate high-quality zero-shot VSC in a streaming setting.
However, its approximately one-second end-to-end latency on consumer hardware remains high for latency-sensitive interaction.
Reducing this latency while preserving holistic conversion quality is therefore a key challenge for making streaming VSC practical.

A second limitation is that existing VSC systems largely require a target reference utterance.
This makes voice specification indirect: users must provide an example voice rather than describe the desired attributes.
In contrast, text-prompted control has become increasingly common in text-to-speech, where natural-language prompts can specify accent, emotion, and fine-grained vocal qualities such as a soft, warm, breathy, or singing-like voice \cite{guo2023prompttts, lacombe-etal-2024-parler-tts, yang2024instructtts, vyas2023audiobox, hu2026qwen3, hu2026voicesculptor, chen2026flexivoice, diwan2025scaling}.
However, these systems synthesize speech from text rather than converting an input utterance, and therefore do not directly solve streaming speech-to-speech VSC.
Recent speech-to-speech conversion work also explores text-guided editing \cite{sheng2025voice, hai2024dreamvoice, yao2024promptvc, qi2025promptevc}, but typically over a single axis such as timbre or emotion.
To our knowledge, no prior system provides text-based voice design and editing for holistic VSC, let alone in a streaming setting.

We present \textbf{StyleStream 2.0}, a fast and controllable streaming VSC framework that addresses both limitations.
StyleStream 2.0 inherits the two-stage architecture of StyleStream 1.0: a Destylizer that extracts content features disentangled from style, and a Stylizer that reconstructs target speech from those content features and a target style.
Because StyleStream 1.0 already established the Destylizer as a strong content-style disentangler, we reuse it without modification and focus our changes on the Stylizer.

To reduce latency, we train a few-step pixel MeanFlow \cite{geng2025mean} Stylizer to replace the multi-step diffusion Stylizer used in StyleStream 1.0.
We further fine-tune this Stylizer with autoregressive feedback inspired by Self Forcing \cite{huang2025self} to improve robustness under chunk-by-chunk inference.
This reduces end-to-end latency to 520 ms on a consumer GPU, a 2$\times$ speedup over StyleStream 1.0, while maintaining conversion quality.

To enable text-based control, we change how style enters the Stylizer.
StyleStream 1.0 conditions the Stylizer on both an unmasked mel-spectrogram context and a style embedding extracted from the target utterance, which leaves style information partially distributed across two conditioning channels.
We remove the mel-spectrogram context so that target style is represented through a single style embedding.
This makes text-based design and editing a problem of predicting or modifying this embedding.
Inspired by the unified image generation and editing model FLUX Kontext \cite{labs2025flux}, we train a unified text-conditioned DiT in the style-embedding space: given a natural-language prompt, the model predicts a target style embedding for \emph{voice design}; given an existing utterance embedding and a text instruction, the same model modifies the embedding along the requested attribute for \emph{voice editing}.

For text-based control, we focus on age, gender, accent, and emotion.
These attributes provide a structured and measurable interface for specifying target voices, while covering the major perceptual factors needed for holistic VSC.
This structured interface supports both prompt-based voice design and attribute-targeted voice editing, and remains largely unexplored for holistic VSC, especially under streaming constraints.

We evaluate StyleStream 2.0 in two settings.
First, for standard target-speech-conditioned VSC, we compare against existing reference-based conversion systems and measure conversion quality, intelligibility, and target style fidelity.
Second, for text-based control, no native holistic VSC baseline exists.
We therefore construct strong cascaded baselines: for voice design, we combine ASR with prompt-controlled TTS; for voice editing, we combine prompt-controlled TTS with reference-based VSC.
These cascades benefit from powerful ASR and TTS systems, yet StyleStream 2.0 achieves stronger results on many metrics while remaining a pure speech-to-speech system that also supports streaming inference.

In summary, our contributions are:
\begin{itemize}
    \item We introduce a faster streaming VSC model by training a few-step pixel MeanFlow Stylizer and fine-tuning it with autoregressive feedback, reducing end-to-end latency to approximately 520 ms on consumer hardware while preserving conversion quality.
    \item We propose a unified text-conditioned style-embedding model for natural-language voice design and attribute-targeted editing over age, gender, accent, and emotion in holistic VSC.
    \item We evaluate StyleStream 2.0 across standard target-speech-conditioned VSC and text-based voice design / editing, showing strong target-style fidelity while remaining competitive in intelligibility, non-edited attribute preservation, and audio quality against both reference-based and cascaded baselines.
\end{itemize}
\section{Related Work}

Prior speech-to-speech conversion work spans several directions, including voice conversion that primarily changes speaker identity while preserving content and other style factors \cite{qian2019autovc, li2023freevc, wang2026diffvc, choi2021neural, baas2023voice, liu2024zero, kaneko2019stargan}, streaming or real-time voice conversion \cite{liu2025rt, yang2024streamvc, ma2026meanvc, wang2024streamvoice}, and separate lines on accent conversion and emotional voice conversion \cite{jia2024convert, bai2025accent, jia2023zero, zhou2022emotion, pan2025clapfm, chou2025zsdevc}. More recent systems have begun to model richer, holistic voice style across timbre, accent, and emotion, including the Vevo family \cite{zhang2025vevo, amphion_v0.2, zhang2025vevo2} and StyleStream~1.0 \cite{liu2026stylestream}. In parallel, prompt-controlled TTS systems enable natural-language specification of target voice characteristics \cite{guo2023prompttts, lacombe-etal-2024-parler-tts, yang2024instructtts, vyas2023audiobox, hu2026qwen3, hu2026voicesculptor, chen2026flexivoice, diwan2025scaling}, with only a few open-source systems explicitly supporting holistic control over timbre, accent, and emotion \cite{wang2025capspeech, hu2026qwen3}. Closer to our setting, a smaller body of work explores text-guided speech-to-speech conversion \cite{sheng2025voice, hai2024dreamvoice, yao2024promptvc, qi2025promptevc}, but these methods typically focus on a single attribute and are evaluated offline. Thus, while prior work shows that speech-to-speech conversion, prompt-based control, and limited text-guided editing are all feasible, no prior system simultaneously provides fast streaming holistic VSC together with unified text-based voice design and editing.

Separately, diffusion and flow-matching models have become strong acoustic generators for speech synthesis and conversion \cite{du2024cosyvoice, du2025cosyvoice, chen2025f5, eskimez2024e2, lee2024ditto, liu2024zero}, but their iterative sampling cost is a bottleneck for streaming applications. Recent few-step generative models reduce this cost by producing high-quality samples in only a few steps, or even one step \cite{frans2024one, song2023consistency, song2023improved, geng2025mean, geng2025improved, lu2026one}. However, fast sampling alone does not guarantee high-quality streaming inference, since short chunks provide limited context and can lead to degraded quality at chunk boundaries. Inspired by Self Forcing for autoregressive video diffusion \cite{huang2025self}, StyleStream~2.0 combines a few-step pixel MeanFlow Stylizer with autoregressive fine-tuning to improve streaming robustness while also enabling controllable holistic voice design and editing.
\section{Method}

StyleStream~2.0 follows the Destylizer-Stylizer framework of StyleStream~1.0 \cite{liu2026stylestream}.
The Destylizer extracts source content features while suppressing speaker timbre, accent, and emotion.
The Stylizer maps these content features and a target style embedding to a mel-spectrogram, which is converted to waveform by a causal Vocos-based vocoder \cite{siuzdak2024vocos}.
The style embedding can be obtained from a reference utterance or from a text-conditioned Voice Controller, allowing the same Stylizer to support reference-based conversion, text-based design, and text-guided editing.

We first recap the Destylizer, then introduce the redesigned Stylizer with pixel MeanFlow training and autoregressive feedback fine-tuning, and finally present the Voice Controller for unified voice design and editing.

\subsection{Destylizer}
\label{sec:destylizer}

We use the Destylizer from StyleStream~1.0.
The Destylizer consists of a frozen HuBERT-Large \cite{hsu2021hubert} encoder followed by six conformer \cite{gulati2020conformer} blocks.
During training, the Destylizer output is passed through a finite scalar quantization (FSQ) \cite{mentzer2023finite} module and then a transformer ASR decoder.
The conformer blocks and the ASR decoder are jointly optimized with a sequence-to-sequence ASR loss.
This training setup encourages the representation to preserve linguistic content, while the compact FSQ bottleneck, with a codebook size of 45, discourages encoding style cues that are not needed for character prediction.
At inference, we discard the FSQ module and ASR decoder, and use the continuous output features immediately before the FSQ layer as content features $f_c \in \mathbb{R}^{D_c \times T}$ at 50~Hz.

\subsection{Stylizer}
\label{sec:stylizer}

\subsubsection{Preliminary: Pixel MeanFlow (pMF)}
\label{sec:pmf_prelim}

Pixel MeanFlow (pMF) \cite{lu2026one} combines JiT-style \cite{li2025back} $x$-prediction with MeanFlow \cite{geng2025mean, geng2025improved} objective for few-step guided generation.
Let $x \sim p_{\mathrm{data}}$ be a clean sample and $\epsilon \sim \mathcal{N}(0,I)$ be Gaussian noise.
We use the linear interpolation path
\begin{equation}
z_t = (1-t)x + t\epsilon, \qquad t \in [0,1],
\label{eq:pmf_path}
\end{equation}
whose instantaneous Flow Matching velocity is
\begin{equation}
v(z_t,t) = \frac{d z_t}{dt} = \epsilon - x .
\label{eq:pmf_fm_velocity}
\end{equation}

JiT motivates predicting the clean sample $x$ rather than noisy or velocity-space quantities, since $x$ lies closer to the data manifold and can be easier for a Transformer to model in high-dimensional spaces.
MeanFlow instead improves sampling efficiency by learning an average velocity over an interval $[r,t]$:
\begin{equation}
u(z_t,r,t)
=
\frac{1}{t-r}
\int_r^t v(z_\tau,\tau)\,d\tau ,
\label{eq:pmf_average_velocity}
\end{equation}
which is related to the instantaneous velocity by the MeanFlow identity
\begin{equation}
v(z_t,t)
=
u(z_t,r,t)
+
(t-r)
\frac{d}{dt}
u(z_t,r,t).
\label{eq:pmf_meanflow_identity}
\end{equation}

pMF combines these two ideas by having the network predict an $x$-like quantity rather than a velocity.
Because pMF uses classifier-free guidance (CFG) distillation, the guidance scale $\omega$ is also provided as an input to the network.
Given condition $c$, the model prediction $\hat{x}_\theta(z_t,r,t,c,\omega)$ is converted to an average velocity by
\begin{equation}
u_\theta(z_t,r,t,c,\omega)
=
\frac{z_t-\hat{x}_\theta(z_t,r,t,c,\omega)}{t}.
\label{eq:pmf_model_average_velocity}
\end{equation}

Let $\varnothing$ denote the null condition.
The conditional and unconditional instantaneous fields are obtained from the degenerate interval $r=t$:
\begin{equation}
v_\theta(z_t,t,c,\omega)
=
u_\theta(z_t,t,t,c,\omega),
\qquad
v_\theta(z_t,t,\varnothing,\omega)
=
u_\theta(z_t,t,t,\varnothing,\omega).
\label{eq:pmf_cond_uncond_velocity}
\end{equation}
Following the pMF CFG distillation formulation, the guided velocity target is
\begin{equation}
v_{\mathrm{cfg}}(z_t,t,c,\omega)
=
(\epsilon - x)
+
\left(1-\frac{1}{\omega}\right)
\left(
v_\theta(z_t,t,c,\omega)
-
v_\theta(z_t,t,\varnothing,\omega)
\right).
\label{eq:pmf_cfg_target}
\end{equation}
When $\omega=1$, the guidance vanishes and the target reduces to the standard Flow Matching velocity.

Using the MeanFlow identity, the instantaneous velocity implied by the predicted average velocity is
\begin{equation}
V_\theta(z_t,r,t,c,\omega)
=
u_\theta(z_t,r,t,c,\omega)
+
(t-r)
\frac{d}{dt}
u_\theta(z_t,r,t,c,\omega),
\label{eq:pmf_implied_velocity}
\end{equation}
where the total derivative is computed with a Jacobian-vector product (JVP).
The final pMF objective is
\begin{equation}
\mathcal{L}_{\mathrm{pMF}}
=
\mathbb{E}_{x,\epsilon,c,r,t,\omega}
\left[
\left\|
V_\theta(z_t,r,t,c,\omega)
-
v_{\mathrm{cfg}}(z_t,t,c,\omega)
\right\|_2^2
\right].
\label{eq:pmf_cfg_loss}
\end{equation}
pMF keeps the network prediction in an $x$-like space while applying supervision in the MeanFlow velocity space needed for few-step guided generation. With CFG distillation, guided sampling requires only one model evaluation per sampling step, further reducing inference cost.

\subsubsection{Stylizer Architecture and Training}

\begin{figure}[t]
    \centering
    \includegraphics[width=\linewidth]{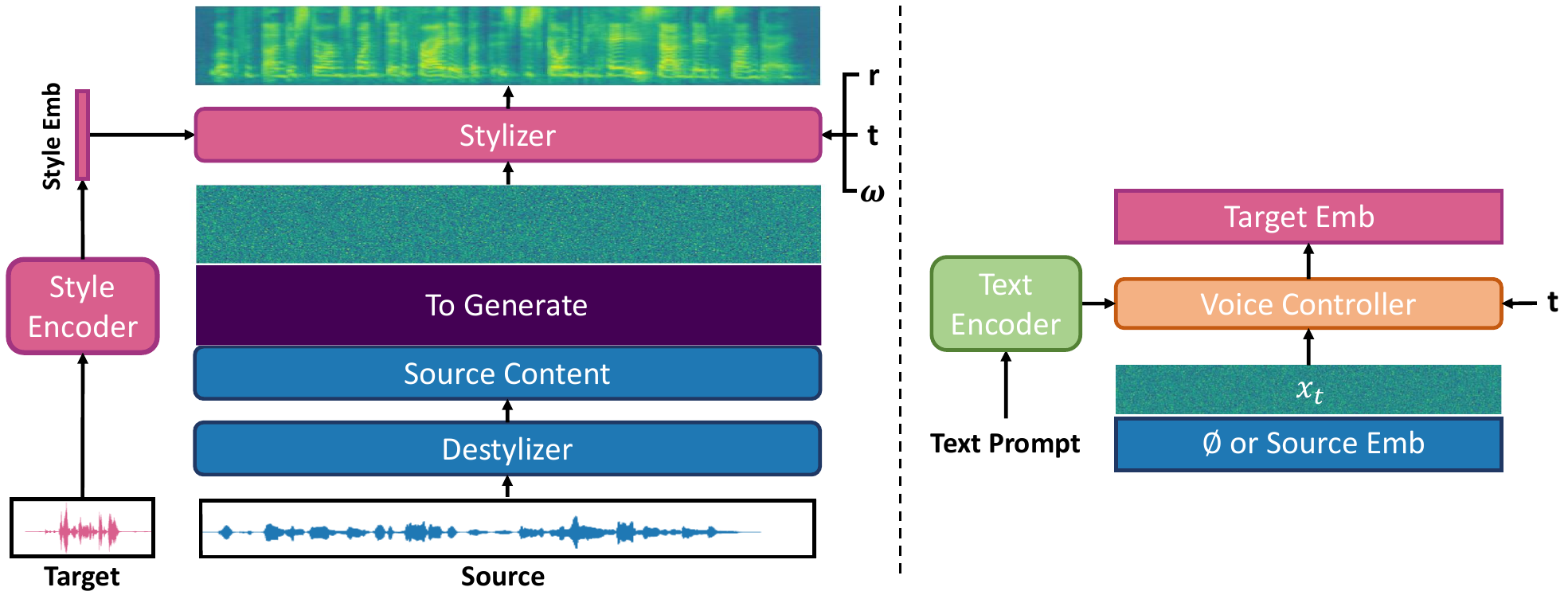}
    \caption{\textbf{Left}: Stylizer architecture. $[r, t]$ is the MeanFlow interval, $\omega$ is the CFG strength. \textbf{Right}: Voice Controller architecture. When used for voice design, the auxiliary input is null $\varnothing$; for voice editing, source style embedding is also part of the input.}
    \label{fig:system}
\end{figure}

\paragraph{Architecture.}
An overview of the Stylizer is in Figure \ref{fig:system}. Our Stylizer is a DiT-based model that follows the overall design of StyleStream~1.0, with two main modifications.
First, we remove the reference mel-spectrogram context used for mel-inpainting in StyleStream~1.0, so target style must be represented entirely by the style embedding.
This makes the style space directly usable for voice design and editing.
Second, because the style embedding is now the only source of target-style information, we replace the original TDNN-based style encoder \cite{jin-etal-2018-tdnn} with a stronger perceiver-resampler encoder \cite{jaegle2021perceiver, alayrac2022flamingo, li2023blip}.

The new style encoder first extracts SSL features from a frozen WavLM \cite{chen2022wavlm} model\footnote{\href{https://huggingface.co/microsoft/wavlm-base-plus}{\texttt{https://huggingface.co/microsoft/wavlm-base-plus}}}.
Let the WavLM hidden states be $H \in \mathbb{R}^{B \times T \times L \times D}$, where $B$ is the batch size, $T$ is the number of frames, $L=12$ is the number of WavLM layers, and $D$ is the hidden dimension.
Instead of reducing $H$ by layer-weighted summation and temporal pooling as in StyleStream~1.0, we concatenate all layer outputs along time to obtain $\tilde{H}\in\mathbb{R}^{B\times (TL)\times D}$.
A perceiver resampler with $N_q$ learnable queries compresses $\tilde{H}$ into a fixed-size representation $S\in\mathbb{R}^{B\times N_q\times D_q}$, through cross-attention to $\tilde{H}$ and self-attention among the queries.
We flatten $S$ and apply an MLP to obtain the style embedding $e_s\in\mathbb{R}^{B\times C}$, where $C$ is the final style embedding dimension.
The style encoder is trained jointly with the Stylizer under the pMF objective, and we compare it with the StyleStream~1.0 style encoder in Section~\ref{sec:ablation}.

\paragraph{Training.}
We train the Stylizer using the pMF objective introduced in Section~\ref{sec:pmf_prelim}.
Here, $x$ is the target mel-spectrogram, $z_t=(1-t)x+t\epsilon$, and the condition is $c=(f_c,e_s)$, where $f_c$ is the Destylizer content representation and $e_s$ is the style embedding.
During training, $f_c$ and $e_s$ are extracted from the same utterance. At inference, $e_s$ may instead come from a different reference utterance or from the Voice Controller (see Section \ref{sec:controller}).

At the DiT input, we concatenate $z_t$, $f_c$, and a past-output channel along the channel dimension.
This past-output channel is zero during offline pretraining and is filled with generated history during autoregressive fine-tuning (Section \ref{sec:ar_finetune}).
The flow times $r$ and $t$, style embedding $e_s$, and guidance scale $\omega$ are embedded as vectors, summed, and injected into the DiT through adaLN-zero \cite{peebles2023scalable}.
The model predicts an $x$-like mel estimate, which is converted into the average velocity $u_\theta(z_t,r,t,c,\omega)$ by Eq.~\eqref{eq:pmf_model_average_velocity} and the implied instantaneous velocity $V_\theta$ by Eq.~\eqref{eq:pmf_implied_velocity}.

For CFG distillation, we randomly sample the guidance scale $\omega$ and replace $c=(f_c,e_s)$ with null condition $\varnothing$ with probability $p_{\mathrm{drop}}$.
When constructing the CFG target velocity, we stop gradients on $v_{\mathrm{cfg}}$ (Eq.~\eqref{eq:pmf_cfg_target}), so the model is updated only through the predicted compound velocity $V_\theta$.
At inference, we provide the desired $\omega$ directly and require only one model evaluation per sampling step.
Additional implementation details are provided in Appendix~\ref{app:implementation}.

\subsubsection{Autoregressive Fine-Tuning}
\label{sec:ar_finetune}

Although pMF reduces the sampling cost, fast sampling alone does not guarantee high-quality streaming generation. When applied chunk by chunk, each generation step only observes a short segment with limited context, making it difficult to produce locally accurate outputs and smooth transitions across chunk boundaries. Inspired by Self Forcing \cite{huang2025self}, we apply autoregressive fine-tuning to adapt the Stylizer to the streaming setting and align training with inference behavior.

During this stage, we freeze the style encoder and update only the Stylizer, so the offline and streaming models remain aligned in the same style embedding space.
Following Self Forcing, which trains generative models on rollouts of their own outputs, we adopt the rollout mechanism while retaining our pMF objective.
For each training utterance, we sample a chunk size and split the utterance into non-overlapping chunks.
The model rolls out chunks sequentially, maintaining a fixed-length window of generated mel frames as past context $m_{\mathrm{past}}$.
Each chunk also receives a fixed-size lookahead window of future content features, which provides limited future linguistic context while preserving the streaming constraint.
The DiT input uses the same layout as offline training, except that the past-output channel is filled with $m_{\mathrm{past}}$ and the current chunk region is zeroed out.
The AR-stage condition is therefore $\tilde{c}=(f_c,e_s,m_{\mathrm{past}})$, with $\omega$ sampled as in offline training.

We apply the pMF loss only on the current chunk:
\begin{equation}
\mathcal{L}_{\mathrm{AR}}
=
\mathbb{E}_{x,\epsilon,\tilde{c},r,t,\omega}
\left[
\left\|
\mathbf{M}
\odot
\left(
V_\theta(z_t,r,t,\tilde{c},\omega)
-
v_{\mathrm{cfg}}(z_t,t,\tilde{c},\omega)
\right)
\right\|_2^2
\right],
\label{eq:ar_pmf_loss}
\end{equation}
where $\mathbf{M}$ selects the current chunk region and $v_{\mathrm{cfg}}$ is defined in Eq.~\eqref{eq:pmf_cfg_target}.
Generated past chunks are detached before being reused, so gradients are computed only through the current chunk loss.
This exposes the Stylizer to its own imperfect history while keeping training efficient.

At inference, we use the same chunk-by-chunk procedure.
The model maintains ring buffers for generated mel frames and past content features.
For each new chunk, the current content features and lookahead are concatenated with the past-content buffer, while the past-mel buffer fills the past-output channel with the current chunk region zeroed out.
After generation, both buffers are updated with the new content and mel frames.

\subsection{Voice Controller: Unified Voice Design and Editing}
\label{sec:controller}

The Voice Controller is a text-conditioned DiT that operates in the style embedding space and supports two tasks: voice design, which generates a new style embedding from a text prompt, and voice editing, which modifies an existing style embedding according to a text instruction.
In this work, both tasks focus on controllable age, gender, accent, and emotion.

As shown in Figure \ref{fig:system}, the Voice Controller is a standard DiT trained with a JiT-style $x$-prediction objective over the style-embedding space.
Given a target style embedding, we add noise to obtain $e_t$, patchify it into a token sequence, and feed these tokens to the DiT.
To unify voice design and editing, we provide an auxiliary embedding sequence with the same patch layout: for voice design, this auxiliary input is a learned null embedding; for voice editing, it is the patchified source style embedding.
The noised target tokens and auxiliary tokens are concatenated along the channel dimension at each patch.
The text prompt is encoded by a frozen Flan-T5 encoder\footnote{\href{http://huggingface.co/google/flan-t5-base}{\texttt{http://huggingface.co/google/flan-t5-base}}} \cite{chung2024scaling} and injected through cross-attention.
For more implementation details, please refer to Appendix~\ref{app:implementation}.

\section{Experimental Setup}
\label{sec:experimental_setup}
\subsection{Datasets}
\subsubsection{Voice Style Conversion}
\label{sec:vsc_data}
\paragraph{Training Data.}
The Destylizer is reused from StyleStream~1.0 without retraining.
We train the Stylizer on the English portion of Emilia \cite{he2024emilia}, which provides roughly 50k hours of in-the-wild English speech covering diverse speakers, accents, and emotions.

\paragraph{Test Data.}
We follow the same construction procedure as StyleStream~1.0.
We sample 300 source utterances from ESD \cite{zhou2022emotional}, GLOBE-test \cite{wang2024globe}, and LibriTTS-test-clean \cite{zen2019libritts}, 
and select 10 target utterances from ESD \cite{zhou2022emotional}, RAVDESS \cite{livingstone2018ryerson}, GLOBE-test, and L2-ARCTIC \cite{zhao2018l2arctic}, 
covering five emotions (happy, angry, sad, fearful, calm) and five accents (British, American, Indian, Arabic, Mandarin).
Pairing each source with each target yields 3000 source-target pairs.

\subsubsection{Voice Design and Editing}

\paragraph{Training Data.}
The Voice Controller is trained on a large synthetic and natural corpus derived from CapSpeech~\citep{wang2025capspeech}. We assign four controllable attributes: age, gender, accent, and emotion, using style information from CapSpeech captions, with VoxProfile~\citep{feng2025vox} predictions used as fallbacks when caption labels are missing or ambiguous. To construct single-attribute editing pairs, we use CosyVoice3~\citep{du2025cosyvoice} to align utterances into a shared timbre space for accent and emotion edits: accent-edit pairs differ primarily in accent while sharing emotion, and emotion-edit pairs differ primarily in emotion while sharing accent. For timbre editing, we use natural CapSpeech pairs selected to differ in age, gender, or both while preserving the remaining available labels.
Categorical labels are converted into natural-language supervision through a controlled template renderer, producing prompts for voice design and single-attribute instructions for voice editing. Full details of the label inventories, template construction, and dataset sizes are provided in Appendix~\ref{app:sft_corpus}.

\paragraph{Test Data.}
We evaluate text-based style control on the same fixed set of 300 source utterances used in Section~\ref{sec:vsc_data}. For voice design, each source utterance is paired with ten natural-language style prompts, yielding 3000 test items. The prompts cover timbre-only requests, accent-only requests, emotion-only requests, timbre+accent requests, timbre+emotion requests, and all-attribute requests. For voice editing, each source utterance is paired with ten single-axis edit instructions, yielding another 3000 test items: two age edits, two gender edits, three accent edits, and three emotion edits. The target value is sampled to differ from the source label on the edited axis, and the instruction mentions only the requested change, so evaluation measures both edit success and preservation of non-edited attributes. Test prompt rendering and target sampling details are given in Appendix~\ref{app:eval_manifests}.

\subsection{Baselines}

\paragraph{Voice Style Conversion.}
For reference-conditioned VSC, we compare against CosyVoice~2.0~\citep{du2024cosyvoice}, CosyVoice~3.0~\citep{du2025cosyvoice}, Vevo~\citep{zhang2025vevo}, Vevo~2.0~\citep{zhang2025vevo2}, and StyleStream~1.0~\citep{liu2026stylestream}. We evaluate CosyVoice in its speech-to-speech mode and run Vevo~2.0 as an ASR+TTS cascade using Whisper-large-v3~\citep{radford2023robust}, since Vevo~2.0 is architecturally a TTS system. StyleStream~2.0 is evaluated in both offline and streaming modes. Sampling, NFE, CFG, chunk-size, and streaming-buffer details are given in Appendix~\ref{app:eval_systems}.

\paragraph{Voice Design and Editing.}
For text-based voice design, we compare against ASR+prompt-TTS cascades using CapTTS~\citep{wang2025capspeech} and Qwen3-TTS VoiceDesign~\citep{hu2026qwen3}. For text-based voice editing, no native holistic text-editing baseline exists, so we use CapTTS as a text-to-reference adapter for Vevo~\citep{zhang2025vevo} and Vevo~2.0~\citep{zhang2025vevo2}. The generated reference is routed as either the timbre or style reference depending on the edited attribute. Full routing details are provided in Appendix~\ref{app:eval_systems}.

\subsection{Metrics}

\paragraph{Voice Style Conversion.}
Following StyleStream~1.0~\citep{liu2026stylestream}, we evaluate reference-conditioned VSC with WER for intelligibility and embedding-based target-style similarity: S-SIM for speaker similarity, A-SIM~\citep{zuluaga2023commonaccent} for accent similarity, and E-SIM~\citep{ma2024emotion2vec} for emotion similarity. Subjective evaluation reports NMOS for naturalness and S/A/E-SMOS for speaker, accent, and emotion similarity to the target reference.

\paragraph{Voice Design and Editing.}
For text-based control, we report WER and target-style accuracy using VoxProfile~\citep{feng2025vox} classifiers for age, gender, accent, and emotion, denoted Age-acc, G-acc, A-acc, and E-acc. For voice editing, we additionally measure preservation of non-edited speaker, accent, and emotion attributes using source-output S-SIM, A-SIM, and E-SIM, denoted as S-pres, A-pres, and E-pres. Subjective evaluation reports NMOS for naturalness, SC-MOS for prompt or instruction consistency, and SP-MOS for preservation in editing. Objective metric details and MOS protocols are provided in Appendices~\ref{app:objective_metrics} and~\ref{app:subjective_protocol}.

\section{Results}
\begin{table*}[t]
\centering
\small
\caption{Voice style conversion results (SS denotes StyleStream). The best results are shown in \textbf{bold}, and the second-best results are \underline{underlined}. Ground-truth NMOS is reported as an upper bound and excluded from ranking.}
\label{tab:main_vsc_results}
\resizebox{\textwidth}{!}{
\begin{tabular}{lccccccccc}
\toprule
\textbf{Model} & \textbf{NFE} $\downarrow$ & \textbf{WER ($\%$)} $\downarrow$ & \textbf{S-SIM} $\uparrow$ & \textbf{A-SIM} $\uparrow$ & \textbf{E-SIM} $\uparrow$ & \textbf{NMOS} $\uparrow$ & \textbf{S-SMOS} $\uparrow$ & \textbf{A-SMOS} $\uparrow$ & \textbf{E-SMOS} $\uparrow$ \\
\midrule
Ground Truth & --- & 4.42 & --- & --- & --- & 3.72{\scriptsize\textcolor{gray}{$\pm$.12}} & --- & --- & --- \\
Qwen3-TTS & --- & 4.76 & 0.927 & 0.833 & 0.941 & --- & --- & --- & --- \\
\midrule
CosyVoice~2.0 & 10 & 8.07 & 0.814 & 0.484 & 0.674 & \textbf{3.54}{\scriptsize\textcolor{gray}{$\pm$.12}} & 3.64{\scriptsize\textcolor{gray}{$\pm$.13}} & 3.25{\scriptsize\textcolor{gray}{$\pm$.15}} & 2.61{\scriptsize\textcolor{gray}{$\pm$.15}} \\
CosyVoice~3.0 & 10 & \underline{6.56} & 0.832 & 0.508 & 0.668 & \underline{3.48}{\scriptsize\textcolor{gray}{$\pm$.13}} & 3.78{\scriptsize\textcolor{gray}{$\pm$.12}} & 3.21{\scriptsize\textcolor{gray}{$\pm$.14}} & 2.80{\scriptsize\textcolor{gray}{$\pm$.15}} \\
Vevo & 32 & 13.33 & 0.853 & 0.659 & 0.745 & 3.42{\scriptsize\textcolor{gray}{$\pm$.13}} & 4.00{\scriptsize\textcolor{gray}{$\pm$.14}} & 3.94{\scriptsize\textcolor{gray}{$\pm$.13}} & 3.63{\scriptsize\textcolor{gray}{$\pm$.13}} \\
Vevo~2.0 & 32 & \textbf{5.79} & 0.843 & 0.676 & 0.723 & 3.44{\scriptsize\textcolor{gray}{$\pm$.13}} & 4.06{\scriptsize\textcolor{gray}{$\pm$.13}} & 3.98{\scriptsize\textcolor{gray}{$\pm$.13}} & 3.11{\scriptsize\textcolor{gray}{$\pm$.15}} \\
SS~1.0 (offline) & 16 & 7.47 & \underline{0.869} & \underline{0.699} & \textbf{0.808} & 3.43{\scriptsize\textcolor{gray}{$\pm$.14}} & \underline{4.23}{\scriptsize\textcolor{gray}{$\pm$.10}} & \textbf{4.08}{\scriptsize\textcolor{gray}{$\pm$.11}} & \underline{4.30}{\scriptsize\textcolor{gray}{$\pm$.10}} \\
SS~2.0 (offline) & 4 & 7.00 & \textbf{0.879} & \textbf{0.705} & \underline{0.774} & 3.45{\scriptsize\textcolor{gray}{$\pm$.13}} & \textbf{4.32}{\scriptsize\textcolor{gray}{$\pm$.10}} & \underline{4.04}{\scriptsize\textcolor{gray}{$\pm$.11}} & \textbf{4.35}{\scriptsize\textcolor{gray}{$\pm$.09}} \\
\midrule
SS~1.0 (streaming) & 12 & 11.41 & 0.842 & 0.656 & 0.771 & 3.37{\scriptsize\textcolor{gray}{$\pm$.13}} & 4.10{\scriptsize\textcolor{gray}{$\pm$.14}} & 3.76{\scriptsize\textcolor{gray}{$\pm$.12}} & 4.24{\scriptsize\textcolor{gray}{$\pm$.10}} \\
SS~2.0 (streaming) & 4 & 12.02 & 0.866 & 0.684 & 0.765 & 3.40{\scriptsize\textcolor{gray}{$\pm$.13}} & 4.18{\scriptsize\textcolor{gray}{$\pm$.14}} & 3.89{\scriptsize\textcolor{gray}{$\pm$.13}} & 4.27{\scriptsize\textcolor{gray}{$\pm$.10}} \\
\bottomrule
\end{tabular}}
\end{table*}
\subsection{Voice Style Conversion}
\label{sec:vsc_results}

Table~\ref{tab:main_vsc_results} reports the main reference-conditioned VSC results in both offline and streaming settings. 
In the offline setting, StyleStream~2.0 achieves the strongest overall target-style similarity, 
ranking first on S-SIM, A-SIM, S-SMOS, and E-SMOS, and improving over the previous state-of-the-art StyleStream~1.0 on most style metrics. 
NMOS is statistically comparable across systems. 
Notably, StyleStream~2.0 achieves these results with only 4 sampling steps, compared with 16 for StyleStream~1.0, 
10 for the CosyVoice variants, and 32 for Vevo and Vevo~2.0. 
For intelligibility, Vevo~2.0 obtains the lowest WER, which is expected because it is effectively a strong cascaded ASR and voice-cloning TTS pipeline. 
However, its target-style similarity remains lower than StyleStream~2.0 across all metrics, while StyleStream~2.0 remains competitive in WER.

In the streaming setting, StyleStream~2.0 matches the intelligibility of StyleStream~1.0 while improving target-style similarity on all metrics except for a marginal gap in E-SIM. 
It also uses only 4 sampling steps (versus 12 for the StyleStream~1.0 streaming variant) and reduces the chunk size from 600~ms to 300~ms, enabling lower-latency streaming inference.

\subsection{Streaming Analysis}

\label{sec:streaming_analysis}

Table~\ref{tab:streaming_chunk} reports streaming quality and per-chunk processing time across chunk sizes from 160~ms to 600~ms, comparing StyleStream~1.0 and StyleStream~2.0 on the same RTX~4060 GPU under the same evaluation protocol. For the streaming configuration reported in Table \ref{tab:main_vsc_results}, we use a 300 ms chunk size and a fixed 80 ms lookahead. Together with the measured per-chunk processing time of 140.7 ms, this gives an end-to-end latency of 520.7 ms, which is 2x faster than StyleStream 1.0.
For both systems, all metrics improve monotonically with chunk size, since longer context lets the model better capture accent and emotion and smooths transitions across chunk boundaries.
At every chunk size, StyleStream~2.0 dominates StyleStream~1.0 on nearly every metric.
StyleStream~2.0 also runs in real time across the entire sweep on RTX 4060, with per-chunk processing time well below the chunk duration, whereas StyleStream~1.0 only meets the real-time requirement at the largest 600~ms chunk and then only with negligible headroom (598~ms processing within a 600~ms budget).

\begin{table}[t]
\centering
\small
\setlength{\tabcolsep}{4pt}
\caption{Streaming chunk-size analysis on RTX~4060. We compare StyleStream~1.0 (SS~1.0, 12 NFE) and StyleStream~2.0 (SS~2.0, 4 NFE) across chunk sizes from 160~ms to 600~ms, reporting per-chunk processing time and quality. Best per chunk-size group is in \textbf{bold}.}
\label{tab:streaming_chunk}
\begin{tabular}{ccccccc}
\toprule
\textbf{Chunk (ms)} & \textbf{Model} & \textbf{Proc (ms)} $\downarrow$ & \textbf{WER ($\%$)} $\downarrow$ & \textbf{S-SIM} $\uparrow$ & \textbf{A-SIM} $\uparrow$ & \textbf{E-SIM} $\uparrow$ \\
\midrule
160 & SS~1.0 & 500.1 & 16.59 & 0.821 & 0.624 & 0.724 \\
160 & SS~2.0 & \textbf{130.5} & \textbf{14.02} & \textbf{0.860} & \textbf{0.669} & \textbf{0.763} \\
\addlinespace
200 & SS~1.0 & 528.8 & 15.25 & 0.825 & 0.632 & 0.732 \\
200 & SS~2.0 & \textbf{132.0} & \textbf{13.22} & \textbf{0.863} & \textbf{0.675} & \textbf{0.764} \\
\addlinespace
300 & SS~1.0 & 559.3 & 13.34 & 0.833 & 0.644 & 0.748 \\
300 & SS~2.0 & \textbf{140.7} & \textbf{12.03} & \textbf{0.866} & \textbf{0.684} & \textbf{0.765} \\
\addlinespace
400 & SS~1.0 & 575.8 & 12.21 & 0.835 & 0.647 & 0.761 \\
400 & SS~2.0 & \textbf{145.9} & \textbf{11.56} & \textbf{0.868} & \textbf{0.687} & \textbf{0.766} \\
\addlinespace
500 & SS~1.0 & 589.5 & 11.62 & 0.837 & 0.651 & 0.764 \\
500 & SS~2.0 & \textbf{150.2} & \textbf{11.12} & \textbf{0.869} & \textbf{0.688} & \textbf{0.766} \\
\addlinespace
600 & SS~1.0 & 598.4 & 11.45 & 0.842 & 0.656 & \textbf{0.772} \\
600 & SS~2.0 & \textbf{157.0} & \textbf{10.87} & \textbf{0.870} & \textbf{0.691} & 0.766 \\
\bottomrule
\end{tabular}
\end{table}

\begin{table*}[t]
\centering
\small
\caption{Voice design results. WER and VoxProfile accuracies are percentages. We use Age-acc, G-acc, A-acc, and E-acc to denote age, gender, accent and emotion classification accuracy. NMOS measures naturalness and SC-MOS measures style consistency with the prompt.}
\label{tab:main_design_results}
\begin{tabular}{lrrrrrrr}
\toprule
& & \multicolumn{4}{c}{Objective} & \multicolumn{2}{c}{Subjective} \\
\cmidrule(lr){3-6} \cmidrule(lr){7-8}
\textbf{Model} 
& \textbf{WER} $\downarrow$ 
& \textbf{Age-acc} $\uparrow$ 
& \textbf{G-acc} $\uparrow$ 
& \textbf{A-acc} $\uparrow$ 
& \textbf{E-acc} $\uparrow$ 
& \textbf{NMOS} $\uparrow$ 
& \textbf{SC-MOS} $\uparrow$ \\
\midrule
CapTTS 
& 7.36 & 41.11 & 94.33 & \underline{52.00} & 29.47 
& 3.27{\scriptsize\textcolor{gray}{$\pm$.13}} 
& 3.54{\scriptsize\textcolor{gray}{$\pm$.13}} \\
Qwen3-TTS 
& \textbf{4.86} & 44.33 & 92.39 & 14.73 & \textbf{36.33} 
& \underline{3.35}{\scriptsize\textcolor{gray}{$\pm$.14}} 
& 2.97{\scriptsize\textcolor{gray}{$\pm$.15}} \\
SS~2.0 (streaming) 
& 10.42 & \underline{67.94} & \underline{98.94} & 46.73 & 30.13 
& 3.23{\scriptsize\textcolor{gray}{$\pm$.13}} 
& \underline{3.78}{\scriptsize\textcolor{gray}{$\pm$.12}} \\
SS~2.0 (offline) 
& \underline{6.89} & \textbf{70.67} & \textbf{99.44} & \textbf{52.73} & \underline{32.20} 
& \textbf{3.45}{\scriptsize\textcolor{gray}{$\pm$.12}} 
& \textbf{3.96}{\scriptsize\textcolor{gray}{$\pm$.10}} \\
\bottomrule
\end{tabular}
\end{table*}
\subsection{Voice Design and Editing}
Table~\ref{tab:main_design_results} summarizes the voice design results. Compared with the cascaded ASR+TTS baselines, StyleStream~2.0 achieves the strongest overall style fidelity, obtaining the highest style accuracy for age, gender, and accent. The streaming variant also improves over the baselines on age and gender accuracy. CapTTS remains competitive for accent control, while Qwen3-TTS achieves the best intelligibility and emotion accuracy, but shows much lower accent accuracy.

Table~\ref{tab:main_edit_results} reports the voice editing results.
StyleStream~2.0 offline achieves the highest edited-axis accuracy and the lowest WER, while also receiving the highest naturalness and style-consistency MOS.
The streaming variant remains close in both style accuracy and SC-MOS, showing that StyleStream~2.0 can follow edit instructions reliably across different style attributes in both offline and streaming settings.
In contrast, Vevo and Vevo~2 obtain lower edited-axis accuracy, especially for accent and emotion edits.
Their higher preservation scores therefore should be interpreted together with their lower edit success: in many cases, the source style is preserved because the requested attribute is not changed sufficiently.
This highlights the trade-off between successful attribute editing and preservation of unedited attributes.

\begin{table*}[t]
\centering
\small
\caption{Voice editing results. WER and VoxProfile accuracies are percentages. We use Age-acc, G-acc, A-acc, and E-acc to denote age, gender, accent and emotion classification accuracy. SC-MOS measures instruction consistency, SP-MOS measures preservation of non-edited attributes, and NMOS measures naturalness.}
\label{tab:main_edit_results}
\resizebox{\textwidth}{!}{
\begin{tabular}{lrrrrrrrrrrr}
\toprule
& & \multicolumn{4}{c}{Style Consistency} & \multicolumn{3}{c}{MOS} & \multicolumn{3}{c}{Style Preservation} \\
\cmidrule(lr){3-6} \cmidrule(lr){7-9} \cmidrule(lr){10-12}
\textbf{Model} 
& \textbf{WER} $\downarrow$ 
& \textbf{Age-acc} $\uparrow$ 
& \textbf{G-acc} $\uparrow$ 
& \textbf{A-acc} $\uparrow$ 
& \textbf{E-acc} $\uparrow$ 
& \textbf{SC-MOS} $\uparrow$ 
& \textbf{SP-MOS} $\uparrow$ 
& \textbf{NMOS} $\uparrow$ 
& \textbf{S-pres.} $\uparrow$ 
& \textbf{A-pres.} $\uparrow$ 
& \textbf{E-pres.} $\uparrow$ \\
\midrule
Vevo 
& 12.35 & 28.33 & 94.83 & 16.89 & 6.56
& 2.64{\scriptsize\textcolor{gray}{$\pm$.16}} 
& \textbf{3.64}{\scriptsize\textcolor{gray}{$\pm$.12}}
& \underline{3.44}{\scriptsize\textcolor{gray}{$\pm$.13}} 
& \textbf{0.9092} & \textbf{0.6303} & \textbf{0.8902} \\
Vevo2 
& \underline{7.82} & 24.67 & 65.17 & 6.78 & 7.11
& 2.26{\scriptsize\textcolor{gray}{$\pm$.15}} 
& 3.28{\scriptsize\textcolor{gray}{$\pm$.14}}
& 3.32{\scriptsize\textcolor{gray}{$\pm$.14}} 
& \underline{0.8853} & \underline{0.5948} & 0.8832 \\
SS~2.0 (streaming) 
& 9.62 & \underline{67.17} & \textbf{99.67} & \underline{46.56} & \underline{13.67}
& \underline{3.64}{\scriptsize\textcolor{gray}{$\pm$.13}} 
& 3.50{\scriptsize\textcolor{gray}{$\pm$.13}}
& 3.21{\scriptsize\textcolor{gray}{$\pm$.15}} 
& 0.7644 & 0.4864 & 0.8794 \\
SS~2.0 (offline) 
& \textbf{6.43} & \textbf{71.67} & \underline{99.17} & \textbf{54.33} & \textbf{14.67}
& \textbf{3.76}{\scriptsize\textcolor{gray}{$\pm$.13}} 
& \underline{3.58}{\scriptsize\textcolor{gray}{$\pm$.12}}
& \textbf{3.48}{\scriptsize\textcolor{gray}{$\pm$.13}} 
& 0.7899 & 0.5102 & \underline{0.8851} \\
\bottomrule
\end{tabular}}
\end{table*}

\subsection{Ablation Studies}
\label{sec:ablation}

Table~\ref{tab:ablations} reports two ablations on the key design choices of StyleStream~2.0: the autoregressive fine-tuning (Section \ref{sec:ar_finetune}) and the perceiver-resampler style encoder (Section \ref{sec:stylizer}).
Without self-forcing, the streaming Stylizer loses substantial intelligibility while target-style similarity stays essentially unchanged, indicating that the autoregressive rollout primarily addresses the train-test mismatch of chunk-by-chunk inference for smooth chunk transitions.
Replacing our perceiver-resampler with the StyleStream~1.0 TDNN style encoder under the same offline pMF objective degrades both intelligibility and accent similarity, supporting the need for a stronger style encoder when the embedding alone must carry timbre, accent, and emotion.
More ablations on the effect of NFE and CFG scale are in Appendix~\ref{app:ablations}. Model-size details and Voice Controller text-encoder ablations are in Appendices~\ref{app:implementation} and~\ref{app:abl_text_encoder}.
\begin{table}[t]
\centering
\caption{Ablation studies. \textbf{Left}: Effect of with or without autoregressive fine-tuning (AR-ft) under streaming inference. \textbf{Right}: Different style encoder effect under offline inference.}
\label{tab:ablations}
\begin{minipage}[t]{0.48\linewidth}
\centering
\resizebox{\linewidth}{!}{
\begin{tabular}{lcccc}
\toprule
\textbf{Setting} & \textbf{WER ($\%$)} $\downarrow$ & \textbf{S-SIM} $\uparrow$ & \textbf{A-SIM} $\uparrow$ & \textbf{E-SIM} $\uparrow$ \\
\midrule
w/ AR-ft  & \textbf{12.03} & 0.866 & 0.684 & 0.765 \\
w/o AR-ft & 19.54 & 0.866 & 0.676 & 0.767 \\
\bottomrule
\end{tabular}}
\end{minipage}
\hfill
\begin{minipage}[t]{0.48\linewidth}
\centering
\resizebox{\linewidth}{!}{
\begin{tabular}{lcccc}
\toprule
\textbf{Style Encoder} & \textbf{WER ($\%$)} $\downarrow$ & \textbf{S-SIM} $\uparrow$ & \textbf{A-SIM} $\uparrow$ & \textbf{E-SIM} $\uparrow$ \\
\midrule
SS~1.0  & 8.04 & 0.823 & 0.633 & 0.765 \\
SS~2.0  & \textbf{7.00} & \textbf{0.879} & \textbf{0.705} & \textbf{0.774} \\
\bottomrule
\end{tabular}}
\end{minipage}
\end{table}

\section{Conclusion}

We introduced StyleStream~2.0, a fast and controllable streaming VSC system. By combining a few-step pixel MeanFlow Stylizer with autoregressive fine-tuning, it reduces streaming latency while preserving holistic style conversion quality. A unified text-conditioned controller further enables natural-language voice design and single-attribute voice editing in the same style-embedding space. Across reference-conditioned VSC, voice design, and editing, StyleStream~2.0 improves target-style fidelity while remaining competitive in intelligibility and preservation.

\begin{ack}


Supported by the IARPA ARTS program, Robert E. \& Beverly A. Brooks Endowed Professorship at EECS, UC Berkeley.
\end{ack}

\bibliographystyle{plainnat}
\bibliography{references}


\appendix

\section{Extended Data Generation and Evaluation Details}
\label{app:text_style_eval}

This appendix gives additional details for the data generation and evaluation protocol used in the StyleStream~2.0 experiments. We describe the text-based style-control corpus, evaluation manifests, baseline routing for reference-conditioned VSC and text-based control, and objective metrics.

\subsection{Training Corpus for Text-Based Style Control}
\label{app:sft_corpus}

The main text summarizes the CapSpeech-derived supervised corpus used to train
the text-conditioned style-control model. The construction proceeds in four
steps: extract candidate attributes from captions and waveforms, normalize and
filter those attributes with confidence rules, form nested utterance
inventories, and finally sample source--target pairs whose text supervision is
rendered from templates. Synthesized portions of the corpus were generated with
CosyVoice3 \citep{du2025cosyvoice} and stored as 24 kHz mono PCM audio.

\paragraph{Caption-attribute extraction.}
CapSpeech captions contain useful natural-language descriptions of speaker
attributes, but the descriptions are not normalized into the evaluation
taxonomy. We therefore first extract explicit caption attributes with
gpt-oss-20b \citep{agarwal2025gpt}. The model is instructed to copy the
caption's own wording for accent and emotion rather than normalizing them, so
that normalization can be performed by a separate deterministic mapping. The
extraction prompt used a structured function call with fields for age, sex,
accent, and emotion:

\begin{quote}
\footnotesize
\begin{verbatim}
You are given a short description of a speaker's voice. Extract ONLY the
attributes the description explicitly states:
  - age (e.g. 'young', 'middle-aged', 'elderly')
  - sex ('male' / 'female')
  - accent (regional / national / dialect speech descriptor)
  - emotion (affective tone)

Rules:
1. Copy the description's own words VERBATIM for accent and emotion. Do not
normalize, translate, bucket, or map them (e.g. if the description says
'Italian accent', return 'Italian', NOT 'Romance'; if it says 'slightly
irritated', return 'slightly irritated', not 'Anger').
2. Return the literal string 'N/A' for any attribute the description does
not mention. Do not infer or guess from unrelated cues.
3. Voice-quality words like 'deep', 'raspy', 'melodic', 'silky', 'husky',
'booming' are NOT emotions -- return 'N/A' for emotion unless an actual
affective tone is stated.
4. Call the extract_speaker_attributes tool exactly once with the four
fields filled in. Output nothing else.
\end{verbatim}
\end{quote}

The function schema required four string fields. Age was copied when stated,
sex was mapped to male/female when explicit, and accent and emotion were copied
verbatim from the caption; any missing attribute was returned as \texttt{N/A}.

\paragraph{Waveform-profile extraction.}
Caption attributes are sparse, especially for accent and emotion, so we also
run VoxProfile \citep{feng2025vox} on each CapSpeech utterance. VoxProfile
provides an age prediction and posterior distributions for sex, accent, and
emotion. These waveform-derived predictions are not used blindly; they serve as
fallback labels only when the caption does not provide a trusted value.

\paragraph{Label normalization and confidence.}
The extracted caption strings are mapped into the training taxonomy with
deterministic normalization rules. Age is bucketed into child [10,15), teen
[15,20), young adult [20,35), middle aged [35,55), and old [55,90). Sex labels
are male/female. The accent vocabulary is North America, British, Scottish,
Northern Irish, Irish, Welsh, Oceania, South African, South Asia, East Asia,
Southeast Asia, Romance, Germanic, Slavic, Semitic, African, and Nordic. The
emotion vocabulary is Happiness, Sadness, Anger, Fear, Disgust, Surprise,
Neutral, Calm, Friendly, and Authoritative; Contempt is dropped from the final
training vocabulary due to lack of samples.

Caption labels that normalize to a first-class bucket take precedence. When a
caption label is missing or cannot be mapped, VoxProfile is used only if it
passes the confidence threshold for that attribute. Sex fallback requires
maximum posterior probability at least 0.99. Accent fallback requires maximum
posterior probability at least 0.8 and margin at least 0.4. Emotion fallback
requires maximum posterior probability at least 0.7 and margin at least 0.35.
Source utterances are further constrained to 2--20 seconds and predicted age in
[10, 90].

\paragraph{Attribute inventories.}
After normalization, we build three nested inventories from CapSpeech
utterances. The broadest inventory, $I_{\mathrm{vc}}$, contains utterances with
confident age and sex labels and is used as a pool of target voice prompts for
voice conversion. The generation inventory, $I_{\mathrm{gen}}$, additionally
requires either an accent or an emotion label. The editing inventory,
$I_{\mathrm{ed}}$, requires all four labels: age, sex, accent, and emotion. The
final inventory sizes are 496,954 rows for $I_{\mathrm{vc}}$, 403,013 rows for
$I_{\mathrm{gen}}$, and 59,613 rows for $I_{\mathrm{ed}}$.

\paragraph{Pair sampling and training tiers.}
Training examples are sampled from the inventories after the label taxonomy and
inventory filters are fixed. For voice design, a source utterance from
$I_{\mathrm{gen}}$ is converted using a target speaker from $I_{\mathrm{vc}}$.
The content is held fixed while the target profile specifies either age, sex,
and accent, or age, sex, and emotion. For accent editing, source and target
utterances share the same emotion but differ in accent; both sides are
re-voiced with a shared timbre so that accent is the intended changing factor.
For emotion editing, source and target utterances share the same accent but
differ in emotion. For timbre editing, pairs are natural recordings selected to
differ in age, sex, or both while preserving the other available labels.
Table~\ref{tab:sft_tiers} summarizes the produced training tiers.

\begin{table}[h]
\centering
\small
\caption{Training tiers for the text-based StyleStream 2.0 corpus.}
\begin{tabular}{lrl}
\toprule
Tier & Samples & Audio source \\
\midrule
Voice design & 2,000,005 & Synthesized \\
Accent editing & 3,000,028 & Synthesized \\
Emotion editing & 2,999,992 & Synthesized \\
Age editing & 1,499,961 & Natural \\
Gender editing & 1,481,314 & Natural \\
Joint age+gender editing & 1,000,001 & Natural \\
\midrule
Total & $\sim$11.98M & \\
\bottomrule
\end{tabular}
\label{tab:sft_tiers}
\end{table}

\paragraph{Template generation and rendering.}
Each sampled example is rendered into a voice-design prompt or edit instruction
using a shared template inventory. We generated this inventory with
gpt-oss-120b \citep{agarwal2025gpt} by providing hand-written seed templates
for each template shape and asking for diverse paraphrases that preserve a
fixed set of placeholders. The template shapes were: all-attribute design,
timbre+accent design, timbre+emotion design, age editing, gender editing, joint
age+gender editing, accent editing, and emotion editing. The total inventory
contains approximately 9000 generated phrasings plus attribute surface-form
expansions such as \textit{American} for North America and \textit{Australian}
or \textit{New Zealand} for Oceania.

The template-generation prompt was:

\begin{quote}
\footnotesize
\begin{verbatim}
You generate caption templates for a text-to-speech dataset.
Shape: <shape name>
Each template MUST contain exactly these placeholders, each used exactly once:
<placeholder list>.
No other placeholders. Do not expand the placeholders into words.
Keep each template to one natural-sounding English sentence, under 25 words.
Vary sentence structure, word choice, and phrasing.

Here are <N> seed templates for reference:
  1. <seed template 1>
  2. <seed template 2>
  ...

Produce <K> NEW templates, distinct from the seeds and from each other.
Output a single JSON array of <K> strings, and nothing else.
\end{verbatim}
\end{quote}

For example, an all-attribute design template required the placeholders
\texttt{\{age\}}, \texttt{\{gender\}}, \texttt{\{accent\}}, and
\texttt{\{emotion\}}, while an accent-editing instruction required only
\texttt{\{tgt\_accent\}}. Generated templates were filtered to require every
specified placeholder exactly once, reject unknown placeholders, and limit the
length of each sentence.

\subsection{Text-Based Evaluation Manifests}
\label{app:eval_manifests}

The main text describes the 300-source, 3000-design-item, and 3000-edit-item
evaluation setup. The source set contains 50 utterances from ESD, 100 from
GLOBE, and 150 from LibriTTS. Below we report the source-label distribution and
target-sampling details that are not shown in the main text.

\paragraph{Evaluation task structure.}
All text-based evaluations use the same fixed 300 source utterances as the reference-conditioned VSC evaluation. The voice-design manifest contains 3000 items, formed by pairing each source with ten style prompts. The voice-editing manifest also contains 3000 items, formed by pairing each source with ten single-axis edit instructions: two age edits, two gender edits, three accent edits, and three emotion edits. For each edit, the target value is sampled to differ from the source label on the edited axis, and the instruction text mentions only the edited attribute.
Evaluation prompts and edit instructions are rendered with the same controlled template renderer used for training.

\paragraph{Voice design mix.}
Table~\ref{tab:design_mix} gives the fixed prompt mix. Timbre denotes the joint
age+gender target.

\begin{table}[h]
\centering
\small
\caption{Voice design prompt mix for evaluation.}
\begin{tabular}{lrr}
\toprule
Design prompt type & Target axes & Count \\
\midrule
Timbre & age, gender & 600 \\
Accent & accent & 600 \\
Emotion & emotion & 600 \\
Timbre+accent & age, gender, accent & 300 \\
Timbre+emotion & age, gender, emotion & 300 \\
All attributes & age, gender, accent, emotion & 600 \\
\bottomrule
\end{tabular}
\label{tab:design_mix}
\end{table}

\paragraph{Voice editing mix.}
The edit manifest contains 600 age edits, 600 gender edits, 900 accent edits,
and 900 emotion edits. The target value is always sampled to differ from the
source label on the edited axis. The instruction text mentions only the edited
attribute; it does not contain explicit ``preserve'' clauses.

\subsection{Systems and Baseline Generation}
\label{app:eval_systems}

The main text lists the compared systems. Here we give the routing details that
are easy to omit from the compact experimental setup.

\paragraph{Reference-conditioned VSC systems.}
For reference-conditioned VSC, we compare against CosyVoice~2.0~\citep{du2024cosyvoice}, CosyVoice~3.0~\citep{du2025cosyvoice}, Vevo~\citep{zhang2025vevo}, Vevo~2.0~\citep{zhang2025vevo2}, and StyleStream~1.0~\citep{liu2026stylestream}. We use the CosyVoice~2.0 and CosyVoice~3.0 flow-matching speech-to-speech usage. Vevo~2.0 is run as an ASR+TTS cascade: Whisper-large-v3~\citep{radford2023robust} transcribes the source and target utterances, and the transcripts together with the target waveform are fed to Vevo~2.0 to synthesize the output. All systems are evaluated with the standard Euler sampler. Baseline NFEs follow each system's recommended setting when applicable: 32 for Vevo and Vevo~2.0, 10 for the CosyVoice variants, and 16 for StyleStream~1.0 offline.

Qwen3-TTS~\citep{hu2026qwen3} is additionally evaluated as an offline Whisper-large-v3 + voice-cloning TTS cascade, using the source transcript and target reference speech.

\paragraph{Streaming inference settings.}
StyleStream~2.0 uses 4 function evaluations (NFE) and a fixed CFG guidance scale $\omega=3$ in both offline and streaming modes. The main streaming configuration uses a 300~ms chunk size and an 80~ms lookahead. During streaming inference, we maintain two 3-second ring buffers: a content ring buffer and a past self-generated mel-context ring buffer. For streaming baselines, we use the largest NFE that satisfies the real-time per-chunk budget on RTX~4060; this gives 12 NFE for StyleStream~1.0 streaming at a 600~ms chunk size.

\paragraph{Vocoder and timing.}
We use the same causal Vocos-based streaming vocoder as StyleStream~1.0~\citep{liu2026stylestream,siuzdak2024vocos}.
Processing times include mel generation and vocoder inference through waveform output; the vocoder contributes approximately 3.139~ms per 300~ms chunk, already included in the reported 140.7~ms processing time.

\paragraph{Design systems.}
For CapTTS \citep{wang2025capspeech} and Qwen3-TTS VoiceDesign
\citep{hu2026qwen3}, the source utterance is first transcribed with Whisper-large-v3
\citep{radford2023robust}; the baseline receives this ASR transcript plus the
design prompt. The known source transcript is reserved for WER evaluation.

\paragraph{Editing systems.}
The editing baselines are Vevo \citep{zhang2025vevo} and Vevo2
\citep{zhang2025vevo2}, using the Amphion implementations \citep{amphion_v0.2}.
For each edit row, CapTTS first generates a target reference WAV using the
post-edit target labels rendered through the same template family and a random
Whisper transcript from the source transcript pool. This reference provides the
requested target style or timbre to Vevo/Vevo2.

For Vevo, we use the Vevo-Style for all text-based editing. For age
or gender edits, the CapTTS reference is used as the timbre reference and the
original source is used as the style reference. For accent or emotion edits, the
original source is used as the timbre reference and the CapTTS reference is used
as the style reference.

For Vevo2, the source utterance is transcribed
with Whisper to form $T_{\mathrm{src}}$, and the CapTTS reference is transcribed
with Whisper to form $T_{\mathrm{tgt}}$. For age or gender edits, Vevo2 receives
the source style reference with $T_{\mathrm{src}}$ and the CapTTS timbre
reference. For accent or emotion edits, Vevo2 receives the CapTTS style
reference with $T_{\mathrm{tgt}}$ and the source timbre reference. In all cases,
the target text for the output is $T_{\mathrm{src}}$.

\subsection{Objective Evaluation Details}
\label{app:objective_metrics}

\paragraph{Similarity models.}
S-SIM is computed with Resemblyzer\footnote{\url{https://github.com/resemble-ai/Resemblyzer}}, A-SIM with the CommonAccent ECAPA model\footnote{\url{https://huggingface.co/Jzuluaga/accent-id-commonaccentecapa}}~\citep{zuluaga2023commonaccent}, and E-SIM with emotion2vec\footnote{\url{https://github.com/ddlBoJack/emotion2vec}}~\citep{ma2024emotion2vec}. For reference-conditioned VSC, similarities are computed between the converted utterance and the target reference. For editing, similarities are computed between the source utterance and the generated edit output.

\paragraph{VoxProfile classifiers.}
The main text defines the attribute accuracies. The VoxProfile
\citep{feng2025vox} checkpoints used for scoring are
\verb|tiantiaf/wavlm-large-age-sex| for age/sex,
\verb|tiantiaf/wavlm-large-narrow-accent| for accent, and
\verb|tiantiaf/whisper-large-v3-msp-podcast-emotion| for emotion. Generated
audio is downmixed to mono, resampled to 16 kHz, and truncated to 15 seconds
before inference. Age predictions are collapsed into young, middle-aged.

\paragraph{WER and similarity.}
WER is computed by transcribing generated audio with Whisper-large-v3~\citep{radford2023robust} and
comparing against the known source transcript, not against the Whisper transcript
used to drive cascaded baselines. For editing, S-SIM, A-SIM, and E-SIM are
computed between the source utterance and the generated edit output. Each
similarity is split according to whether the corresponding factor should be
preserved or changed by the instruction; preservation scores are better when
higher, while change scores are better when lower.

\subsection{Subjective Evaluation Protocol}
\label{app:subjective_protocol}

\paragraph{Recruitment and screening.}
All subjective evaluations are deployed on Prolific\footnote{\url{https://www.prolific.com}}.
We recruit 40 participants. We restrict participation to listeners whose first language is English and who report fluent English proficiency, with a Prolific approval rating of at least 95\%.
Submissions completed faster than the expected minimum task duration are automatically rejected to filter low-effort responses, and listeners are encouraged to take the survey on a laptop or desktop with headphones for the best audio experience. 

\begin{figure}[t]
\centering
\begin{minipage}[t]{0.48\textwidth}
\centering
\includegraphics[width=\linewidth]{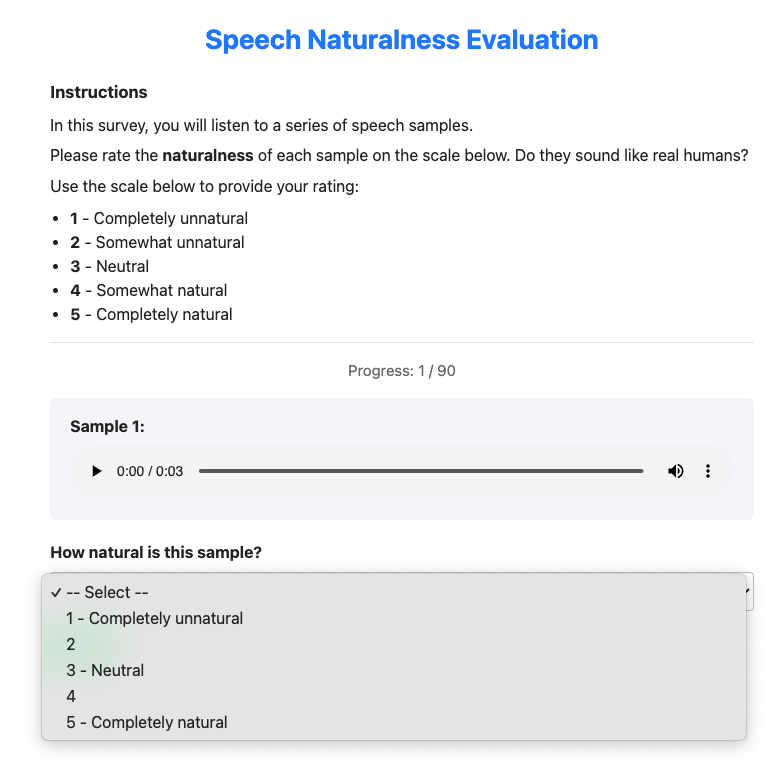}\\[2pt]
{\small (a) NMOS}
\end{minipage}\hfill
\begin{minipage}[t]{0.48\textwidth}
\centering
\includegraphics[width=\linewidth]{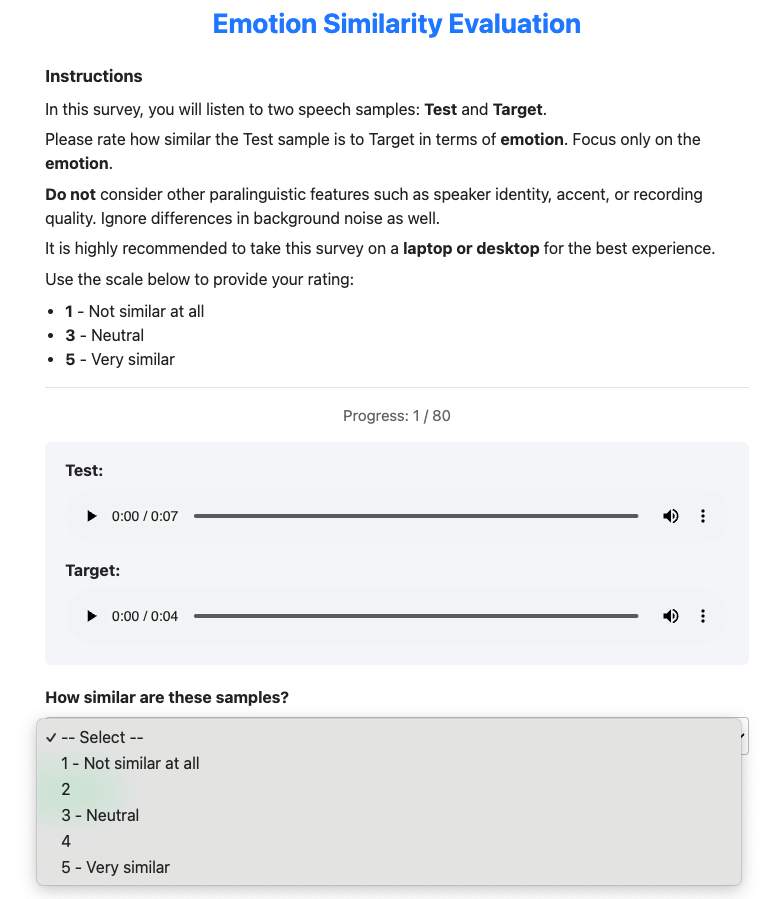}\\[2pt]
{\small (b) E-SMOS}
\end{minipage}
\caption{Naturalness MOS (NMOS) and Emotion SMOS (E-SMOS) evaluation interfaces. The S-SMOS and A-SMOS surveys are analogous to E-SMOS, with the emotion target replaced by speaker identity or accent and the listener instructed to ignore the other style axes.}
\label{fig:nmos_esmos}
\end{figure}

\begin{figure}[t]
\centering
\begin{minipage}[t]{0.48\textwidth}
\centering
\includegraphics[width=\linewidth]{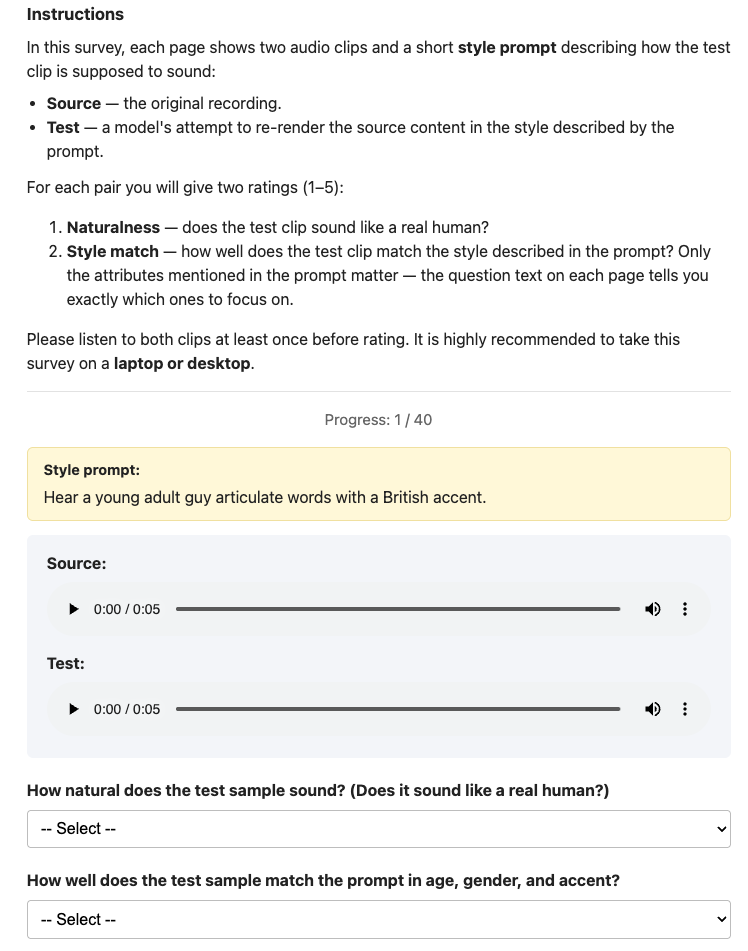}\\[2pt]
{\small (a) Voice design (VD)}
\end{minipage}\hfill
\begin{minipage}[t]{0.48\textwidth}
\centering
\includegraphics[width=\linewidth]{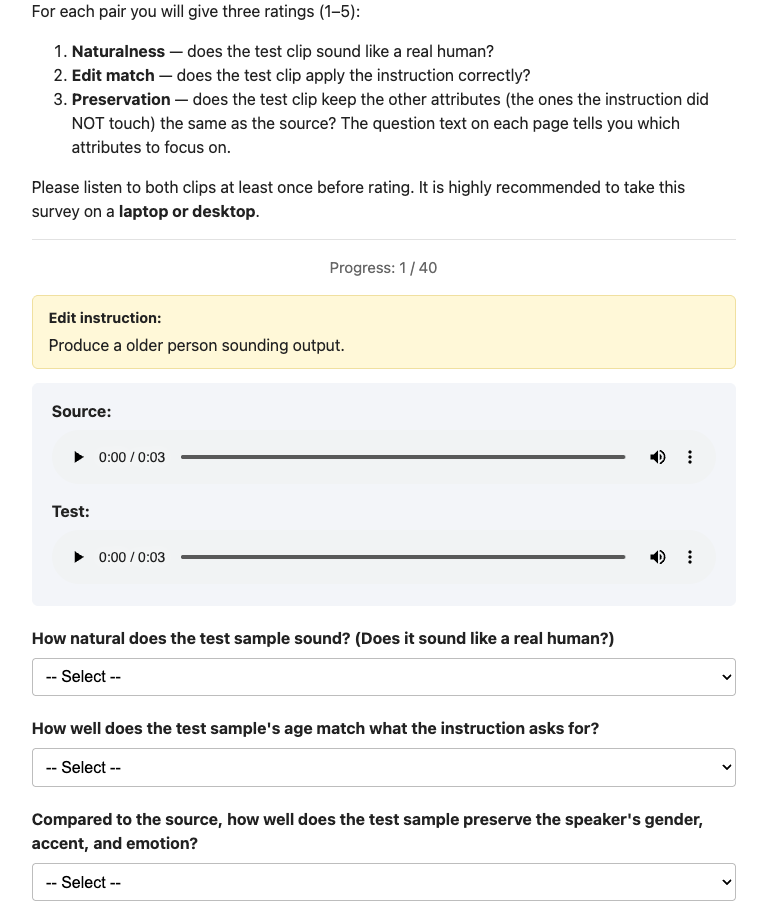}\\[2pt]
{\small (b) Voice editing (VE)}
\end{minipage}
\caption{Voice-design (VD) and voice-editing (VE) evaluation interfaces. Each item pairs a source and a test audio sample with a natural-language style prompt (VD) or edit instruction (VE), and asks for naturalness plus task-specific match (and preservation, for VE) ratings.}
\label{fig:vd_ve}
\end{figure}

\paragraph{Survey types.}
We deploy four types of subjective surveys, all using a 5-point integer scale: NMOS for naturalness (90 items per survey); SMOS for style similarity, in three variants: S-SMOS (speaker), A-SMOS (accent), E-SMOS (emotion), 
with 80 paired items per variant; voice-design (VD) for text-prompted voice design (40 items per survey); and voice-editing (VE) for instruction-based voice editing (40 items per survey).
Representative survey interfaces are shown in Figures~\ref{fig:nmos_esmos} and~\ref{fig:vd_ve}. 

We compensate participants at a rate of at least \$20 per hour.
The NMOS studies take approximately 10 minutes on average, corresponding to \$3.50 per participant.
The SMOS studies take approximately 20 minutes on average, corresponding to \$7.00 per participant.
The voice design and voice editing studies take approximately 10 minutes on average, corresponding to \$3.50 per participant.

\paragraph{NMOS.}
Listeners rate the naturalness of a single test audio sample on a 5-point scale (1: completely unnatural, 5: completely natural), guided by the question ``Do they sound like real humans?''.

\paragraph{SMOS (speaker, accent, emotion).}
For each SMOS survey, listeners are presented with a \textit{Test} and a \textit{Target} audio pair, and rate similarity along the indicated style axis on a 5-point scale (1: not similar at all, 5: very similar).
S-SMOS rates speaker timbre, A-SMOS rates accent, and E-SMOS rates emotion.
Listeners are explicitly instructed to focus only on the indicated axis and to ignore the other style axes as well as recording-quality differences.

\paragraph{Voice design (VD).}
Each item pairs a \textit{Source} recording with a \textit{Test} audio (the model's attempt to re-render the source content in the prompted style), accompanied by a natural-language style prompt (e.g., ``Hear a young adult guy articulate words with a British accent.'').
Listeners give two 5-point ratings: naturalness, and style match against the prompt.
The question text on each page lists exactly which prompt-mentioned attributes the listener should focus on for the style-match rating.

\paragraph{Voice editing (VE).}
Each item pairs a \textit{Source} recording with a \textit{Test} audio (the source after the requested edit), accompanied by a natural-language edit instruction (e.g., ``Produce an older person sounding output.'').
Listeners give three 5-point ratings: naturalness, edit match (does the test apply the instruction correctly?), and preservation (does the test keep the non-edited attributes the same as the source?).
The question text on each page lists which attribute is being edited and which attributes should be preserved.

\section{Implementation Details}
\label{app:implementation}

This section reports the full architecture and training configurations of all StyleStream~2.0 components.
The Destylizer is reused from StyleStream~1.0 without modification.

\subsection{Stylizer}
\label{app:impl_stylizer}

\paragraph{Architecture.}
The Stylizer is a 16-layer DiT with hidden dimension 768, FFN expansion ratio 4 (FFN dimension 3072), and 12 attention heads.
StyleStream~1.0 uses the same 16-layer Stylizer DiT backbone.
We use ALiBi positional encoding \cite{press2021train} and adaLN-Zero conditioning \cite{peebles2023scalable} driven by the summed embedding of $(r, t, e_s, \omega)$.
At the input, $z_t$ (100 mel bins), $f_c$ (768-dimensional content features), and the past-mel-context channel (100 mel bins, zeroed during offline training) are concatenated along the channel dimension and projected to the hidden size.
The output is a 100-channel $x$-prediction of the target mel-spectrogram, from which the average velocity $u_\theta$ and the implied instantaneous velocity $V_\theta$ are computed as in \S\ref{sec:stylizer}.

\paragraph{Style encoder.}
The style encoder uses a frozen \texttt{microsoft/wavlm-base-plus} backbone, with the outputs of all 12 transformer layers concatenated along the time axis as in \S\ref{sec:stylizer}.
The perceiver-resampler stacks 4 blocks; each block first cross-attends from $N_q = 8$ learnable queries (query dimension 128) to the WavLM features, then self-attends among the queries, with 4 attention heads in both stages.
The flattened pooled representation of size $N_q \cdot 128 = 1024$ is mapped through an MLP to the output style embedding $e_s \in \mathbb{R}^{C}$ with $C = 768$.
Including the frozen speech backbone, the StyleStream~2.0 style encoder has 97.5M parameters compared with 128.4M for StyleStream~1.0.

\paragraph{Mel-spectrogram and audio configuration.}
Audio is sampled at 16~kHz and converted to 100-bin mel-spectrograms at a 50~Hz frame rate, matching the Destylizer output rate.
Training uses 6-second segments.

\paragraph{Offline pMF training.}
The Stylizer is trained from scratch on Emilia-EN with the pMF objective.
For each training utterance we sample $r$ and $t$ with $r = t$ on 75\% of training samples (a pure flow-matching step, where the loss reduces to direct supervision of the instantaneous velocity) and $r < t$ on the remaining 25\% (a MeanFlow average-velocity step, with the total derivative in $V_\theta$ computed via JVP).
The diffusion time is sampled as logitnormal distribution $t = \sigma(\ell)$ with $\ell \sim \mathcal{N}(P_{\mathrm{mean}}, P_{\mathrm{std}}^2)$ and $P_{\mathrm{mean}} = P_{\mathrm{std}} = 0.8$, and replaced with $t \sim \mathcal{U}[0, 1]$ on 10\% of training samples. Here $\sigma(\cdot)$ is the sigmoid function. 
The guidance scale is sampled as $\omega = (1 + \omega_{\max})^u$ with $u \sim \mathcal{U}[0, 1]$ and $\omega_{\max} = 7$, equivalent to log-uniform sampling on $[1, 1 + \omega_{\max}]$.
We drop the condition $c = (f_c, e_s)$ with probability $p_{\mathrm{drop}} = 0.2$, replacing it with $\varnothing$.
We optimize with AdamW at peak learning rate $10^{-4}$, weight decay $10^{-6}$, 2000 warmup steps and cosine decay to $10^{-6}$, in bf16-mixed precision, for 1M steps with a global batch size of 256 on 8 H100 GPUs.

\paragraph{Autoregressive fine-tuning.}
We initialize from the offline pMF checkpoint and fine-tune with the Self Forcing chunked rollout while keeping the pMF objective; the FM / MeanFlow split is shifted to 90\%~/~10\% so that most fine-tuning steps are pure flow-matching.
The $t$ sampling, $\omega$ sampling, and conditioning dropout follow the offline stage.
Each rollout uses one ODE step per chunk with a fixed rollout-time guidance scale $\omega_{\mathrm{rollout}} = 3$.
For each batch we randomly sample the chunk size $C \in \{8, \ldots, 30\}$ frames (160 to 600~ms at 50~Hz), and use a 3-second prefix and a 3-second past-context window (150 frames each at 50~Hz).
The training loss is computed only on the current chunk region.
We optimize with AdamW at peak learning rate $10^{-4}$, weight decay $10^{-6}$, 1000 warmup steps and cosine decay to $10^{-6}$, in bf16-mixed precision, for 400k steps with a global batch size of 64 on 32 H100 GPUs.

\subsection{Voice Controller}
\label{app:impl_voicecontrol}

\paragraph{Architecture.}
The Voice Controller is a standard 16-layer DiT with hidden dimension 1024, FFN expansion ratio 4 (FFN dimension 4096), and 16 attention heads, using RoPE positional encoding \cite{su2024roformer}.
It is an additional module for text-based design/editing and is not involved in reference-conditioned VSC.
The text prompt is encoded by a frozen Flan-T5 base\footnote{\href{http://huggingface.co/google/flan-t5-base}{\texttt{http://huggingface.co/google/flan-t5-base}}} and injected via cross-attention.
The model operates on style embeddings of dimension $C = 768$.
We patchify the noised embedding $e_t$ with patch size $P = 32$ into $C/P = 24$ tokens of dimension 32, and concatenate them channel-wise with an auxiliary input of the same shape (a learned null embedding for voice design, the patchified source style embedding for voice editing) before projecting to the hidden dimension.
Style embeddings are normalized using the global training-set mean and standard deviation during training, and de-normalized at inference.

\paragraph{Training.}
We train the Voice Controller jointly on voice design and voice editing with the x-prediction flow-matching objective, splitting each mini-batch evenly between the two tasks.
The diffusion time is sampled from the same logit-normal distribution as the Stylizer (\S\ref{app:impl_stylizer}), $t = \sigma(\ell)$ with $\ell \sim \mathcal{N}(P_{\mathrm{mean}}, P_{\mathrm{std}}^2)$ and $P_{\mathrm{mean}} = P_{\mathrm{std}} = 0.8$.
We apply standard CFG dropout on the text prompt with probability $p_{\mathrm{drop}} = 0.2$.
We optimize with AdamW at peak learning rate $10^{-4}$, weight decay $10^{-6}$, 1000 warmup steps and cosine decay to $3\times10^{-5}$, in bf16-mixed precision, for 500k steps with a global batch size of 1024 on 8 H100 GPUs.

\paragraph{Inference.}
We sample the target style embedding with the standard two-pass CFG sampler, using NFE = 64 and CFG guidance scale $\omega = 4$.
The generated embedding is de-normalized with the stored training-set statistics before being passed to the Stylizer as $e_s$. The controller prepares this embedding once before streaming, and it is reused across chunks.

\section{Additional Ablation Studies}
\label{app:ablations}

We report two further inference-time ablations of the offline StyleStream~2.0 model: the CFG guidance scale $\omega$ and the number of pMF sampling steps (NFE).
Both use the same evaluation protocol as Table~\ref{tab:main_vsc_results}. We additionally ablate Voice Controller text conditioning below.

\subsection{CFG Guidance Scale}
\label{app:abl_cfg}

Table~\ref{tab:abl_cfg} sweeps $\omega \in \{1, 2, 3, 4, 5, 6\}$ at NFE${}=4$.
Removing guidance ($\omega=1$) hurts every metric and pushes WER from around 7\% up to 9.53\%.
For $\omega \ge 2$, WER stays in a narrow band; the similarity metrics show shallow optima with S-SIM and A-SIM peaking near $\omega=2$ and E-SIM near $\omega=4$, while S-SIM declines steadily as $\omega$ grows.
We set $\omega = 3$ as the default operating point, which sits inside the high-similarity band on every axis while keeping WER below 7.1\%.

\begin{table}[h]
\centering
\small
\setlength{\tabcolsep}{8pt}
\caption{CFG guidance-scale ablation on the offline StyleStream~2.0 model at NFE${}=4$.}
\label{tab:abl_cfg}
\begin{tabular}{ccccc}
\toprule
$\omega$ & WER ($\%$) $\downarrow$ & S-SIM $\uparrow$ & A-SIM $\uparrow$ & E-SIM $\uparrow$ \\
\midrule
1 & 9.53 & 0.8806 & 0.6852 & 0.7462 \\
2 & 7.19 & 0.8835 & 0.7071 & 0.7711 \\
3 & 7.03 & 0.8791 & 0.7055 & 0.7751 \\
4 & 6.98 & 0.8741 & 0.7000 & 0.7766 \\
5 & 6.97 & 0.8694 & 0.6929 & 0.7747 \\
6 & 7.00 & 0.8653 & 0.6870 & 0.7722 \\
\bottomrule
\end{tabular}
\end{table}

\subsection{Number of Sampling Steps (NFE)}
\label{app:abl_steps}

Table~\ref{tab:abl_steps} sweeps NFE $\in \{1, 2, 4, 8, 16\}$ at $\omega = 3$.
Thanks to pMF training, the gap between one-step and many-step inference is small: the one-step model already reaches WER 7.06\% (on par with every higher-NFE setting) 
and similarity scores within roughly 0.01 of the saturated values, so the one-step and few-step variants are themselves strong generators.
All similarity metrics are essentially saturated from NFE${}=2$ onwards.
We adopt NFE${}=4$ in the main paper as a robust operating point, while the one-step and two-step variants remain attractive options for latency-sensitive deployments.

\begin{table}[h]
\centering
\small
\setlength{\tabcolsep}{8pt}
\caption{NFE ablation on the offline StyleStream~2.0 model at $\omega = 3$.}
\label{tab:abl_steps}
\begin{tabular}{ccccc}
\toprule
NFE & WER ($\%$) $\downarrow$ & S-SIM $\uparrow$ & A-SIM $\uparrow$ & E-SIM $\uparrow$ \\
\midrule
1  & 7.06 & 0.8755 & 0.6976 & 0.7689 \\
2  & 7.05 & 0.8791 & 0.7052 & 0.7740 \\
4  & 7.02 & 0.8787 & 0.7058 & 0.7744 \\
8  & 7.06 & 0.8785 & 0.7047 & 0.7768 \\
16 & 7.09 & 0.8791 & 0.7056 & 0.7756 \\
\bottomrule
\end{tabular}
\end{table}

\subsection{Voice Controller Text Conditioning}
\label{app:abl_text_encoder}

We train four Voice Controllers for 150k steps with identical DiT architecture, training data, and inference settings, comparing randomly initialized trainable token embeddings (using the Flan-T5 tokenizer) with frozen Flan-T5-small/base/XL encoders.
These controlled runs use a shorter training budget than the 500k-step controller in the main results.
Tables~\ref{tab:abl_text_design} and~\ref{tab:abl_text_edit} report the voice-design and editing results.

Pretrained Flan-T5 representations substantially improve instruction following: design gender accuracy rises from 81.17\% to 97.61--98.33\%, and editing gender accuracy from 60.00\% to 92.17--97.33\%, while WER remains similar.
Larger encoders give mixed, attribute-dependent gains; Flan-T5-base performs best on several voice-design metrics, supporting our choice of base as a balance between semantic conditioning and model size.

\begin{table}[htbp]
\centering
\small
\setlength{\tabcolsep}{5pt}
\caption{Voice-design text-encoder ablation after 150k controller training steps. WER and attribute accuracies are percentages. Flan-T5 encoders are frozen; trainable embeddings are randomly initialized. \textbf{Bold}/\underline{underline} mark the best/second-best values within this ablation.}
\label{tab:abl_text_design}
\begin{tabular}{lrrrrr}
\toprule
Text conditioning & WER $\downarrow$ & Age-acc $\uparrow$ & G-acc $\uparrow$ & A-acc $\uparrow$ & E-acc $\uparrow$ \\
\midrule
Trainable embeddings & \textbf{6.80} & 59.67 & 81.17 & 45.60 & 29.47 \\
Flan-T5-small & 7.02 & \underline{61.06} & \underline{97.94} & 47.87 & 33.60 \\
Flan-T5-base & 7.05 & \textbf{64.56} & \textbf{98.33} & \textbf{49.20} & \textbf{39.93} \\
Flan-T5-XL & \underline{6.91} & 58.94 & 97.61 & \underline{49.13} & \underline{38.40} \\
\bottomrule
\end{tabular}
\end{table}

\begin{table}[htbp]
\centering
\small
\setlength{\tabcolsep}{3pt}
\caption{Voice-editing text-encoder ablation after 150k controller training steps, under the same conditions as Table~\ref{tab:abl_text_design}. WER and accuracies are percentages; preservation scores are similarities to the source. Best/second-best values include ties.}
\label{tab:abl_text_edit}
\begin{tabular}{lrrrrrrrr}
\toprule
Text conditioning & WER $\downarrow$ & Age-acc $\uparrow$ & G-acc $\uparrow$ & A-acc $\uparrow$ & E-acc $\uparrow$ & S-pres. $\uparrow$ & A-pres. $\uparrow$ & E-pres. $\uparrow$ \\
\midrule
Trainable embeddings & \textbf{6.77} & \underline{60.00} & 60.00 & 43.56 & 12.56 & 0.5873 & \underline{0.4480} & \textbf{0.8723} \\
Flan-T5-small & \underline{6.86} & 54.17 & \underline{97.00} & 47.44 & 16.11 & 0.5795 & 0.4350 & 0.8689 \\
Flan-T5-base & \underline{6.86} & \textbf{61.50} & 92.17 & \underline{49.00} & \underline{22.89} & \underline{0.5960} & \textbf{0.4535} & \underline{0.8721} \\
Flan-T5-XL & 6.95 & 55.00 & \textbf{97.33} & \textbf{49.44} & \textbf{24.33} & \textbf{0.5992} & 0.4433 & 0.8718 \\
\bottomrule
\end{tabular}
\end{table}

\section{Limitations}
\label{app:limitations}

StyleStream~2.0 reduces end-to-end streaming latency by roughly $2\times$ over StyleStream~1.0, but the resulting 520~ms still leaves room for improvement in tightly interactive use cases.
On the controllability side, our timbre interface covers age and gender; finer-grained vocal qualities such as breathiness, brightness, and vocal effort would broaden the design and editing interface and are a natural next step.
Emotion editing, while ahead of all baselines, still trails the corresponding timbre and accent results, and editing introduces some drift on non-targeted attributes; further reducing these gaps is left to future work.

\section{Broader Impacts}
\label{app:broader_impacts}

StyleStream~2.0 enables low-latency, controllable voice style conversion and text-conditioned voice design and editing.
Beneficial applications include accessibility tools, voice-based privacy and anonymization, accent localization for media, language-learning aids, and creative audio production.

Like other high-quality generative voice systems, StyleStream~2.0 also carries risks of misuse, particularly for voice impersonation, fraud, or synthetic disinformation.
Our training data comes entirely from existing publicly released speech corpora, with no additional scraping of identifiable individuals.
As safeguards for responsible release, we plan to release model weights under a license that prohibits impersonation and other malicious applications, 
accompanied by a model card documenting training data, intended uses, and known biases, and we support continued research on synthetic-speech detection and audio watermarking.

\section{Licenses and Terms for Existing Assets}
\label{app:asset_licenses}

Table~\ref{tab:asset_licenses} summarizes the licenses or stated terms of use for the main external datasets, pretrained models, toolkits, and evaluation models used in this work. We use all assets for non-commercial academic research and cite the corresponding papers or repositories. For assets with non-commercial, research-only, or responsible-use terms, we restrict our use to the permitted research and evaluation setting.

\begin{table}[h]
\centering
\small
\setlength{\tabcolsep}{4pt}
\caption{Licenses and terms for existing assets used in this work.}
\label{tab:asset_licenses}
\resizebox{\textwidth}{!}{
\begin{tabular}{lll}
\toprule
\textbf{Asset} & \textbf{License / terms} & \textbf{Source} \\
\midrule
Emilia / Emilia-EN & CC BY-NC 4.0; audio copyright remains with original owners & \url{https://huggingface.co/datasets/amphion/Emilia} \\
CapSpeech & CC BY-NC 4.0 & \url{https://github.com/WangHelin1997/CapSpeech} \\
ESD & Free for research use only & \url{https://hltsingapore.github.io/ESD/download.html} \\
GLOBE & CC0 1.0 & \url{https://huggingface.co/datasets/MushanW/GLOBE} \\
LibriTTS & CC BY 4.0 & \url{https://www.openslr.org/60/} \\
RAVDESS & CC BY-NC-SA 4.0 & \url{https://zenodo.org/records/1188976} \\
L2-ARCTIC & CC BY-NC 4.0 & \url{https://psi.engr.tamu.edu/l2-arctic-corpus/} \\
\midrule
HuBERT / fairseq & Apache 2.0 checkpoint; MIT code & \url{https://huggingface.co/facebook/hubert-large-ls960-ft} \\
WavLM & CC BY-SA 3.0 / UniSpeech license terms & \url{https://huggingface.co/microsoft/wavlm-large} \\
Flan-T5 base & Apache 2.0 & \url{https://huggingface.co/google/flan-t5-base} \\
Whisper / Whisper-large-v3 & MIT & \url{https://github.com/openai/whisper} \\
gpt-oss-20b / gpt-oss-120b & Apache 2.0 plus usage policy & \url{https://openai.com/index/gpt-oss-model-card/} \\
\midrule
CosyVoice2 / CosyVoice3 & Apache 2.0 & \url{https://github.com/FunAudioLLM/CosyVoice} \\
Vevo checkpoint & CC BY-NC 4.0 & \url{https://huggingface.co/amphion/Vevo} \\
Vevo2 checkpoint & CC BY-NC-ND 4.0 & \url{https://huggingface.co/RMSnow/Vevo2} \\
Amphion toolkit & MIT & \url{https://github.com/open-mmlab/Amphion} \\
\midrule
VoxProfile classifiers & Open RAIL; excludes clinical, surveillance, privacy-invasive, and commercial use & \url{https://huggingface.co/tiantiaf} \\
Resemblyzer & Apache 2.0 & \url{https://github.com/resemble-ai/Resemblyzer} \\
CommonAccent ECAPA model & MIT & \url{https://huggingface.co/Jzuluaga/accent-id-commonaccent_ecapa} \\
emotion2vec & Model license as specified by the model card; used for research evaluation & \url{https://huggingface.co/emotion2vec/emotion2vec_plus_large} \\
\bottomrule
\end{tabular}}
\end{table}



\end{document}